\documentclass[sigplan]{acmart}
\renewcommand\footnotetextcopyrightpermission[1]{}

\renewcommand\footnotetextcopyrightpermission[1]{}
\usepackage{times}
\usepackage{mathtools}
\usepackage{enumitem}
\usepackage[normalem]{ulem} 
\usepackage[export]{adjustbox}

\usepackage{titlesec}
\titlespacing*\section{0pt}{6pt plus 0pt minus 0pt}{2pt plus 0pt minus 0pt}
\titlespacing*\subsection{0pt}{6pt plus 0pt minus 0pt}{2pt plus 0pt minus 0pt}
\titlespacing*\subsubsection{0pt}{6pt plus 0pt minus 0pt}{2pt plus 0pt minus 0pt}

\usepackage{algorithm}
\usepackage[noend]{algpseudocode} 
\usepackage{algorithmicx}

\usepackage{etoolbox}
\usepackage{tikz}
\usetikzlibrary{tikzmark}
\usetikzlibrary{calc}

\newcommand{\ALGtikzmarkcolor}{black}
\newcommand{\ALGtikzmarkextraindent}{4pt}
\newcommand{\ALGtikzmarkverticaloffsetstart}{-.5ex}
\newcommand{\ALGtikzmarkverticaloffsetend}{-.5ex}
\makeatletter
\newcounter{ALG@tikzmark@tempcnta}

\newcommand\ALG@tikzmark@start{%
	\global\let\ALG@tikzmark@last\ALG@tikzmark@starttext%
	\expandafter\edef\csname ALG@tikzmark@\theALG@nested\endcsname{\theALG@tikzmark@tempcnta}%
	\tikzmark{ALG@tikzmark@start@\csname ALG@tikzmark@\theALG@nested\endcsname}%
	\addtocounter{ALG@tikzmark@tempcnta}{1}%
}

\def\ALG@tikzmark@starttext{start}
\newcommand\ALG@tikzmark@end{%
	\ifx\ALG@tikzmark@last\ALG@tikzmark@starttext
	\else
	\tikzmark{ALG@tikzmark@end@\csname ALG@tikzmark@\theALG@nested\endcsname}%
	\tikz[overlay,remember picture] \draw[\ALGtikzmarkcolor] let \p{S}=($(pic cs:ALG@tikzmark@start@\csname ALG@tikzmark@\theALG@nested\endcsname)+(\ALGtikzmarkextraindent,\ALGtikzmarkverticaloffsetstart)$), \p{E}=($(pic cs:ALG@tikzmark@end@\csname ALG@tikzmark@\theALG@nested\endcsname)+(\ALGtikzmarkextraindent,\ALGtikzmarkverticaloffsetend)$) in (\x{S},\y{S})--(\x{S},\y{E});%
	\fi
	\gdef\ALG@tikzmark@last{end}%
}

\apptocmd{\ALG@beginblock}{\ALG@tikzmark@start}{}{\errmessage{failed to patch}}
\pretocmd{\ALG@endblock}{\ALG@tikzmark@end}{}{\errmessage{failed to patch}}

\usepackage{cleveref}

\crefformat{section}{\S#2#1#3} 
\crefformat{subsection}{\S#2#1#3}
\crefformat{subsubsection}{\S#2#1#3}

\newcommand{\zerognn}{\textsc{ZeroGNN}}

\usepackage{listings}
\usepackage{xcolor}

\newcommand{\bren}[1]{[{\color{purple}BR: #1}]}

\definecolor{codegreen}{rgb}{0,0.6,0}
\definecolor{codegray}{rgb}{0.5,0.5,0.5}
\definecolor{codepurple}{rgb}{0.58,0,0.82}
\definecolor{backcolour}{rgb}{0.95,0.95,0.92}

\lstdefinestyle{mystyle}{
  commentstyle=\color{codegreen},
  keywordstyle=\color{magenta},
  numberstyle=\tiny\color{codegray},
  stringstyle=\color{codepurple},
  basicstyle=\ttfamily\footnotesize,
  breakatwhitespace=false,         
  breaklines=true,                 
  captionpos=b,                    
  keepspaces=true,                 
  numbers=left,                    
  numbersep=5pt,                  
  showspaces=false,                
  showstringspaces=false,
  showtabs=false,                  
  tabsize=2
}
\usepackage{balance}

\begin{document}
%

\title[Metadata-Driven Host Overheads in GNN Training]{Understanding and Reducing Metadata-Driven Host Overheads in Sampling-Based GNN Training}

\author{Yidong Gong}
\affiliation{%
  \institution{William \& Mary}
  \city{Williamsburg}
  \state{VA}
  \country{USA}
}

\author{Saima Afrin}
\affiliation{%
  \institution{William \& Mary}
  \city{Williamsburg}
  \state{VA}
  \country{USA}
}

\author{Yuchen Ma}
\affiliation{%
  \institution{William \& Mary}
  \city{Williamsburg}
  \state{VA}
  \country{USA}
}

\author{Guannan Wang}
\affiliation{%
  \institution{William \& Mary}
  \city{Williamsburg}
  \state{VA}
  \country{USA}
}

\author{Bin Ren}
\affiliation{%
  \institution{William \& Mary}
  \city{Williamsburg}
  \state{VA}
  \country{USA}
}

\author{Pradeep Kumar}
\affiliation{%
  \institution{William \& Mary}
  \city{Williamsburg}
  \state{VA}
  \country{USA}
}




%


\begin{abstract}

Modern deep learning workloads increasingly exhibit dynamic, metadata-driven execution, where runtime-generated information determines memory provisioning and kernel launch decisions. In sampling-based graph neural network (GNN) training, this behavior places the CPU on the critical path, introducing persistent host–device orchestration overhead and frequent GPU–CPU synchronization, which dominate end-to-end runtime when GPU computation is small. Existing approaches, including CUDA Graphs and GPU dynamic parallelism, fail to address this problem because the metadata-driven control loop remains host-mediated, and execution structure varies across iterations. We present {\zerognn}, a system that removes the host from the metadata-driven control loop and enables fully GPU-resident execution under dynamic behavior. {\zerognn} keeps runtime metadata on-device, mediates dynamic execution within a fixed launch structure, and provisions a conservative yet tight execution envelope to restore CUDA Graph replayability. Experiments on sampling-based GNN workloads show that {\zerognn} achieves up to 5.28 $\times$ end-to-end speedup, near-100\% GPU execution fraction, and memory efficiency comparable to ideal metadata-informed allocation, while enabling strong multi-GPU scaling by eliminating host-side bottlenecks.

\end{abstract}

\date{}
\maketitle


%


\section{Introduction} \label{sec.intro}
In today’s data-driven applications, deep learning (DL) has become a core technique for both training and inference~\cite{fan2019graph, zitnik2018modeling,hamilton2018embedding,ying2018graph,krahmer2003graph,perera2017recent,rgcn2018,wang2020entity}. To meet the growing computational demand, GPUs are widely deployed as the primary accelerators. Meanwhile, modern DL workloads are no longer purely static: many emerging models introduce additional sub-tasks and runtime-dependent execution behaviors, which significantly alter how DL systems are orchestrated and executed.
A key consequence of this trend is that the CPU plays an increasingly critical role in metadata-driven orchestration. Beyond launching GPU kernels, the CPU participates in control-flow decisions, runtime scheduling, and memory provisioning based on intermediate results of execution. As a result, CPU execution is frequently placed on the performance-critical path, making host-side orchestration a major limiter for end-to-end performance in dynamic DL workloads.

This work focuses on sampling-based Graph Neural Network (GNN) training, a representative and widely adopted class of dynamic deep learning workloads.
GNNs are used to learn from graph-structured data such as social networks, recommendation graphs, and knowledge graphs. Early GNNs often employed full-batch training (e.g., GCN~\cite{gcn17iclr}, GAT~\cite{gat18iclr}, GIN~\cite{xu2019powerful}, and others~\cite{gatedgraph2017,zhang18}), in which the entire graph is processed per iteration. While effective, full-batch execution consumes prohibitive memory and frequently causes out-of-memory failures even for mid-sized graphs.
To scale GNN training, modern systems adopt mini-batch training with neighbor sampling (e.g., GraphSAGE~\cite{graphsage17}). Each iteration first samples a small subgraph and then performs training on that subgraph. This approach greatly reduces memory usage and enables training on much larger graphs under the same GPU memory budget. However, neighbor sampling fundamentally introduces dynamic behavior: sampled subgraphs vary across iterations, leading to runtime-dependent vertex/edge counts and irregular computation patterns.
This dynamic behavior makes optimization challenging. 

Prior work ~\cite{Graphpy} has shown that in sampling-based GNN training, when per-iteration GPU computation becomes small, the end-to-end runtime is dominated by framework overhead, i.e., CPU-side execution such as Python/C/C++ control logic, kernel launches, and runtime orchestration operations. We refer to this CPU-side cost of driving and coordinating GPU execution as \textit{Host–Device Orchestration Overhead(HDOO)}. Importantly, HDOO tends to remain relatively constant and does not shrink proportionally with subgraph size. At the same time, sampling produces much smaller GPU workloads than full-batch training, and thus GPU kernels execute quickly on modern accelerators. Consequently, the bottleneck shifts toward HDOO, reducing overall system efficiency.

Importantly, the challenge goes beyond “CPU overhead exists.” In sampling-based execution, key runtime metadata, such as sampled vertex/edge counts, is produced on the GPU but immediately required to drive follow-up control decisions, including kernel launch configurations, memory allocations, and tensor preparation. This creates frequent GPU $\rightarrow$ CPU metadata round-trips and synchronization points in the critical path, forming a \textit{host-mediated dependency barrier(HMDB)}: the CPU cannot proceed without waiting for metadata, and the GPU cannot execute subsequent steps without CPU decisions. As a result, the system loses overlap opportunities between phases and underutilizes available resources.
This bottleneck also limits scaling. Ideally, data-parallel multi-GPU training~\cite{p3gandhi2021, torch-data-parallel, Horovod, distributed-para} reduces per-GPU computation by splitting the mini-batch across devices, achieving near-linear speedups. However, in sampling-based GNN training, dividing the mini-batch often produces even smaller subgraphs per GPU, further shrinking per-device GPU compute time. Meanwhile, Host–Device Orchestration Overhead (HDOO) includes kernel launches, synchronizations, and metadata-driven control, remains largely unchanged per iteration, and does not decrease with GPU count. Therefore, HDOO becomes a strong-scaling limiter and prevents effective multi-GPU speedups, even though inter-GPU communication for GNNs can be relatively small due to their modest model parameter sizes. 

A natural direction for reducing host overhead is CUDA Graphs, which can remove repeated per-iteration launch and orchestration costs by capturing and replaying a fixed GPU execution graph. CUDA Graphs can be effective when workload behavior is static and predictable. Unfortunately, dynamic, metadata-driven execution renders this approach ineffective: runtime-dependent control flow and dynamic memory requirements break the predictability assumptions of capture-and-replay. As a result, although framework overhead dominance is well recognized, a practical and general solution remains missing for dynamic DL workloads.

We ask:
\textit{How to eliminate Host–Device Orchestration Overhead and remove host-mediated dependency barrier for metadata-driven dynamic workloads, enabling GPU-only execution efficiency when runtime behavior varies across iterations?}

To address this, we propose {\zerognn}, enabling GPU-only execution for dynamic, metadata-driven workloads by:

\noindent $\bullet$
\textit{ Device-Resident Metadata Buffer (DRMB)} eliminates the host-mediated dependency barrier (HMDB) by introducing memory indirection to keep the metadata flow on the GPU, eliminating frequent GPU $\rightarrow$ CPU round-trips and host-side synchronization in the critical path.

\noindent $\bullet$
\textit{Enabling fully device-side launch and provisioning for metadata-driven workloads} by combining Device-Side Launch Mediation (DLM) and a Metadata-Free Dispatcher (MFD): kernel-launch indirection moves per-iteration launch decisions to the GPU for device-side launch, while MFD provisions a safe execution envelope with conservative upper bounds for memory and launch resources, avoiding under-provisioning and eliminating host involvement.

\noindent $\bullet$
\textit{Unlocking CUDA Graph replay under dynamic behavior.}
With HMDB removed and launch/provisioning handled fully on-device, {\zerognn} recovers the key capture/replay conditions and thus unlocks CUDA Graph replay for dynamic, metadata-driven workloads to reduce per-iteration overhead.


Evaluation shows that {\zerognn} shows substantial performance improvements over existing systems. On average, {\zerognn} achieves 5.28 $\times$, 2.92 $\times$, and 2.33 $\times$ end-to-end training runtime speedups compared to DGL~\cite{dgl2019}, GraphPy~\cite{Graphpy}, and CU-DPI, respectively.
When focusing solely on the sampling phase, {\zerognn} achieves even greater acceleration, 17.69 $\times$, 7.41 $\times$, and 12.75 $\times$ speedups over the same baselines.
In terms of memory efficiency, {\zerognn} attains up to 3.41 $\times$ lower memory usage than DGL, and approximately 10.84 $\times$ memory savings compared to the naive maximal allocation strategy(MaxSG). Furthermore, it achieves memory consumption comparable to the optimal dynamic allocation approach(GraphPy~\cite{Graphpy}), which allocates subgraph memory precisely according to runtime metadata.
Moreover, {\zerognn} achieves the highest \textit{GPU execution fraction}, around 100 \%,  among all evaluated frameworks, including DGL and GraphPy, confirming that {\zerognn} is the only one that eliminates host-side bottlenecks. It further preserves strong multi-GPU scaling by eliminating per-worker host orchestration overhead. We also show that {\zerognn} remains effective in large-graph settings, achieving on average 2.59 $\times$ end-to-end speedups.

The remainder of the paper is organized as follows. The background is presented in ~\cref{sec.background}, motivation and overview in ~\cref{sec.analysis}, detailed
design and discussion in ~\cref{sec.arch}, and evaluations in ~\cref{sec.exp}. Other related
works and discussions are presented in ~\cref{sec.related}, and we conclude in ~\cref{sec.conclusion}.

\section{Background} \label{sec.background}

\subsection{GNN Storage Formats.}
In a graph G = (V, E), V and E refer to the vertex/row set and the edges/non-zero elements, respectively. 
We continue to use both the graph and sparse linear algebra terminologies. Specifically, features and computation are referred to as vertex-level and edge-level, while rows, columns, and non-zero elements (NZE) refer to datasets.
The \textit{compressed sparse row} (CSR) format stores NZE in a row sequentially and uses the offset array to point to the start of the row. For a directed graph, its transpose is also stored; where CSR stores the rows consecutively, while \textit{compressed sparse column} (CSC) format stores the columns consecutively. The \textit{degree} of a row is the row-length.





\subsection{Sampling-based GNN iteration.}
\label{sec.back.gnn}

Given a graph G = (V, E), modern large-scale GNN training commonly adopts mini-batch training with neighbor sampling. Each training iteration starts from a mini-batch of labeled source vertices and performs multi-hop sampling to construct a set of sampled subgraphs, followed by feature gathering and GNN computation on the sampled subgraphs.

\noindent
\textbf{Subgraph sampling.}
Given a mini-batch $V_{s}^{1}$, the sampler expands neighbors to generate a destination set 
$V_{d}^{1}$ and the corresponding edges, where the \textit{fan-out} k denotes the number of neighbors sampled per source vertex. For an 
N-layer GNN, sampling is performed for multiple hops; we denote the source and destination vertex sets at hop i as $V_{s}^{i}$ and $V_{d}^{i}$, respectively.  Importantly, the sampled vertex/edge counts ($V_{s}^{i}$, $V_{d}^{i}$, $|E|^{i}$) vary across each iteration, making sampling-based training a representative metadata-driven dynamic workload.

\noindent
\textbf{ID translation (relabeling). }
The sampled subgraphs are initially expressed in the original global ID space, which would lead to prohibitively large tensor footprints (e.g., $|V|\times F$) if used directly. Therefore, systems perform ID translation (relabeling) to remap the active vertices in the sampled subgraph to a compact, contiguous local ID space (while keeping the adjacency sparse) and to construct mapping tables between global IDs and local IDs. This step determines the exact tensor shapes and buffer sizes required by subsequent stages.

\noindent
\textbf{Feature/label Copy. }
After relabeling, features associated with sampled vertices (typically $V_{d}^{N}$) are gathered from the full feature table using the generated ID mapping, and labels for source vertices (typically $V_{d}^{1}$) are also collected. These operations are indexed and irregular, and their working-set sizes depend on the sampled subgraph structure.

\noindent
\textbf{Subgraph GNN Training.} 
Finally, the GNN model executes on the sampled subgraphs. Compared to full-batch training, each iteration operates on significantly smaller graphs, resulting in short GPU kernels whose end-to-end performance can become sensitive to CPU-side orchestration overhead in existing frameworks.

\subsection{CUDA Graphs and predictability assumptions.}
\label{sec.back.cudagraph}
CUDA Graphs can reduce repeated per-iteration CPU overhead by capturing and replaying a fixed GPU execution graph, amortizing kernel launch and orchestration costs. This technique is effective when workload behavior is static and predictable. 
However, sampling-based training introduces runtime-dependent control flow and dynamic memory requirements. In particular, runtime metadata produced during sampling and relabeling directly determines tensor sizes, memory provisioning, and kernel launch parameters in later stages. Such dynamic, metadata-driven execution violates the predictability assumptions required by capture-and-replay, making CUDA Graphs difficult to apply directly in existing frameworks.

\section{Motivation and Design Overview} \label{sec.analysis}
This section characterizes sampling-based GNN training to identify the dominant performance bottlenecks under realistic training configurations. Our goal is to establish key observations that motivate a system-level solution for eliminating framework overhead. Unless otherwise noted, all results are measured on the Reddit dataset using GraphSAGE~\cite{graphsage17} with default hyperparameters following prior works~\cite{graphsage17,cluster-gcn}.

\begin{figure}[b]
 \centering
  \includegraphics[scale= 0.48]{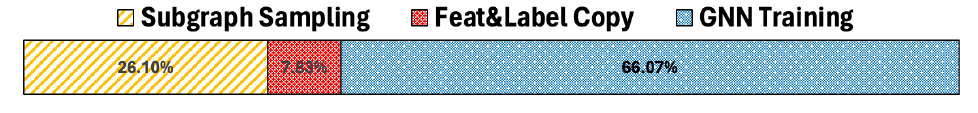}
  \vspace{-12pt}
  \caption{\small Stage-wise Breakdown of End-to-End Training Time (\%)}
  \label{fig-stage}
\end{figure}

\subsection{Workload Characterization: The Real Bottleneck}

\noindent
\textbf{Observation 1.} \textit{End-to-end performance is primarily determined by the sampling and training stages, making optimization of these two stages critical for accelerating sampling-based GNN workloads.}

Fig.~\ref {fig-stage} presents the stage-wise breakdown of the end-to-end runtime for the stages described in \cref{sec.back.gnn}. We observe that stages (a) sampling and (c) training dominate the overall runtime, accounting for 26\% and 66\%, respectively, while stage (b) feature/label copying contributes only 8\%. 




\noindent
\textbf{Observation 2.} \textit{Sampling-based GNN training is frequently HDOO-dominated, leaving significant GPU execution capacity underutilized.}

Prior works~\cite{Graphpy, gnnbench} showed that GNN training leads to HDOO-dominated, where CPU-side orchestration overhead outweighs GPU compute time. In such cases, optimizing GPU kernels alone provides limited end-to-end benefits, and reducing HDOO is necessary to unlock meaningful acceleration.
To quantify whether execution is GPU-bound or HDOO-bound, we measure GPU Execution Fraction, defined as:


{\setlength{\abovedisplayskip}{1pt}
\setlength{\belowdisplayskip}{4pt}
\begin{equation*}
\text{GPU Execution Fraction} = \frac{\text{GPU Time}}{\text{End-to-End Training Runtime}}
\end{equation*}
}

\begin{figure}[b]
  \centering
  \includegraphics[scale= 0.40,center]{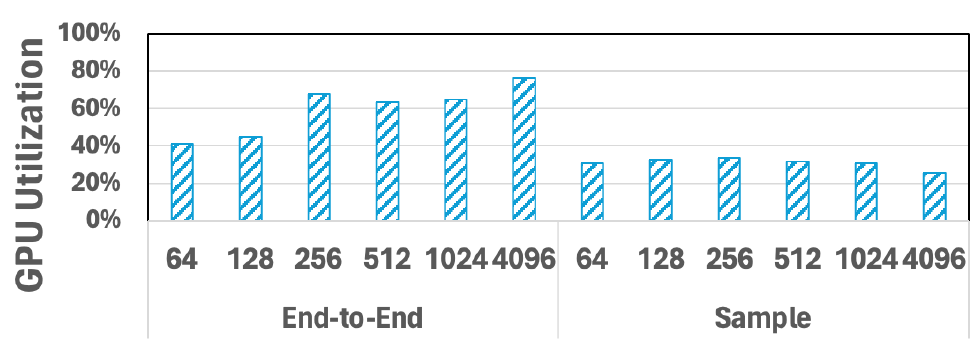}
\vspace{-18pt}
  \caption{\small Illustration of PU Execution Fraction for DGL  GraphSAGE Stages Across Different Batch Sizes
  }
  \label{fig-dgl-util}
\end{figure}

Fig.~\ref{fig-dgl-util} reports GPU execution fraction for GraphSAGE on Reddit across different batch sizes. We find that GPU execution fraction in both the overall pipeline and the training stage remains relatively low when the batch size is smaller than
4096. Specifically, when considering batch size 128, only 45\% of the end-to-end runtime corresponds to active GPU computation, while the remaining 55\% is GPU idle time. 

This low GPU execution fraction indicates that the end-to-end runtime is dominated by HDOO, including CPU-side framework logic, repeated kernel launches, and synchronization required by metadata-driven orchestration.







\noindent
\textbf{Observation 3.} \textit{Conventional framework-level optimizations reduce HDOO constants but Plateau.}

\begin{figure}[t]
  \centering
  \includegraphics[scale= 0.57,center]{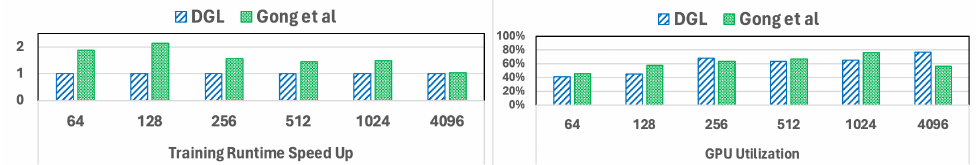}
\vspace{-18pt}
  \caption{\small Illustration of end-to-end training runtime and corresponding GPU Utilization comparison between DGL and Gong et al across Different Batch Sizes
  }
  \label{fig-dgl-graphpy}
\end{figure}

The conventional approach to mitigate HDOO includes:
(i) simplifying framework design by eliminating redundant components, and
(ii) migrating performance-critical modules from Python to C++.
Fig.~\ref{fig-dgl-graphpy} compares DGL with a representative lightweight framework design. Although the lightweight approach~\cite{Graphpy} achieves higher end-to-end speedups (e.g., 1.56 $\times$ at batch size 256), its GPU execution fraction remains comparable to or even lower than DGL (e.g., 63\% vs. 68\% at batch size 256).
This suggests that trimming framework code improves throughput, but the GPU still cannot sustain near-continuous execution due to persistent HDOO and synchronization points.

In summary, sampling-based GNN training frequently operates in an overhead-dominated regime, where reducing GPU kernel time alone is insufficient for improving end-to-end performance. We next show that this bottleneck arises from dynamic metadata that repeatedly forces the host into the execution-critical path.

\subsection{Root Cause: Dynamic Metadata Creates HMDB and Breaks Replayability}

A natural way to reduce repeated host-side launch overhead is CUDA Graphs, which capture and replay a fixed GPU execution graph to amortize per-iteration CPU launch costs. However, CUDA Graphs assume predictable control flow, stable memory addresses, and fixed launch structures. Unfortunately, sampling-based GNN workloads fundamentally violate these assumptions due to dynamic, metadata-driven execution (\cref{sec.back.cudagraph}).
Its runtime metadata is produced on the GPU each iteration and must be repeatedly \textit{materialized as CPU-resident scalars} and consumed by the host control plane to construct subsequent allocations and kernel launches, forcing GPU$\rightarrow$CPU round-trips that insert HMDB and break capture-and-replay.

\begin{figure}[b]
 \centering
  \includegraphics[scale= 0.53]{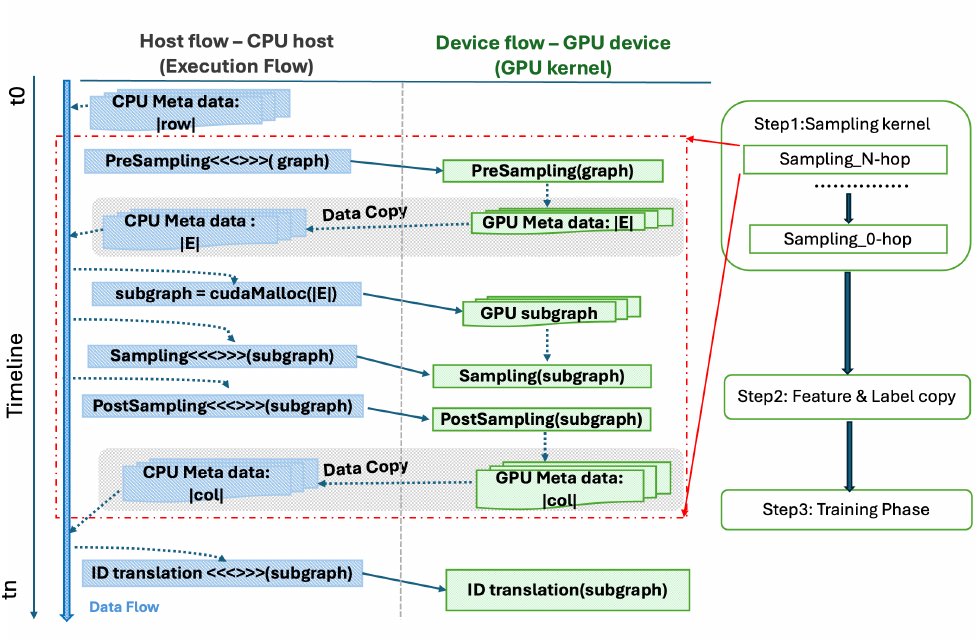}
  \vspace{-12pt}
  \caption{\small Execution Flow: Host-Device Coordination in Sampling-Based GNNs}
  \label{fig-dynamic-data}
\end{figure}

Fig.~\ref{fig-dynamic-data} illustrates this behavior with a representative multi-hop sampling GNN workflow. Within each hop $\ell$, GPU kernels \textit{preSampling(graph)} and \textit{postSampling(subgraph)} generate hop-specific runtime metadata such as the sampled vertex/edge counts $|V_\ell|$ and $|E_\ell|$. These metadata immediately trigger two nested dependency structures.

\noindent
\textbf{(a)	Intra-hop dependency.}
Within the same hop, later GPU execution depends on CPU-resident metadata.
The metadata produced by early kernels in hop $\ell$ must be materialized as CPU scalars because the host uses them to provision downstream buffers and tensors and to configure subsequent kernel launches (e.g., grid sizes and kernel arguments).
This creates an intra-hop “Produce $\rightarrow$ Export $\rightarrow$ Consume $\rightarrow$ Relaunch” control loop.

\noindent
\textbf{(b) Inter-hop Dependency}
Second, across hops, the input of hop $\ell{+}1$ depends on the output of hop $\ell$.
Multi-hop sampling therefore forms a strict dependency chain rather than a flat sequence of independent kernels: each hop produces runtime metadata that determines the frontier and workload of the next hop. After all hops complete, these per-hop metadata are further consumed by downstream stages such as ID translation and GNN training. As a result, runtime metadata drives execution not only within individual sampling hops, but across the entire sampling-to-training pipeline.
This nested dependency structure leads to two fundamental consequences.

\noindent
\textbf{Consequence 1: Dynamic metadata repeatedly introduces a host-mediated dependency barrier (HMDB).}
Because runtime metadata must be transferred to and consumed by the host before subsequent execution decisions can be made, the training pipeline repeatedly incurs GPU$\rightarrow$CPU metadata exchange and the associated host-device synchronization. As a result, the host becomes a required part of the critical path.

\noindent
\textbf{Consequence 2: Nested dependencies exacerbate HDOO and violate CUDA Graph replay assumptions.}
Because downstream allocation and kernel-launch decisions continue to depend on runtime metadata consumed by the host, execution can no longer progress as a purely GPU-driven asynchronous flow. Instead, kernel launches must repeatedly wait for host-side decisions, which serializes launch progression and reduces effective overlap and parallel execution across stages. These synchronization and orchestration costs remain on the execution path and do not shrink proportionally with GPU kernel time, so they increasingly dominate end-to-end runtime. More fundamentally, because these host-side decisions vary with runtime metadata across iterations, the resulting execution no longer satisfies the stable control flow, launch structure, and memory behavior required for CUDA Graph capture and replay.

In short, this host-mediated execution pattern misaligned with modern GPU programming interfaces, which are moving toward fully asynchronous, GPU-resident progress (e.g., beyond asynchronous kernels, even allocation/free are becoming asynchronous via \textit{cudaMallocAsync}/\textit{cudaFreeAsync}). This motivates {\zerognn}: to recover replayability, the system must keep metadata on-device and transform metadata-driven execution into a replayable GPU-resident flow.

\subsection{Overview of ZeroGNN}

\begin{figure}[b]
 \centering
  \includegraphics[scale= 0.53]{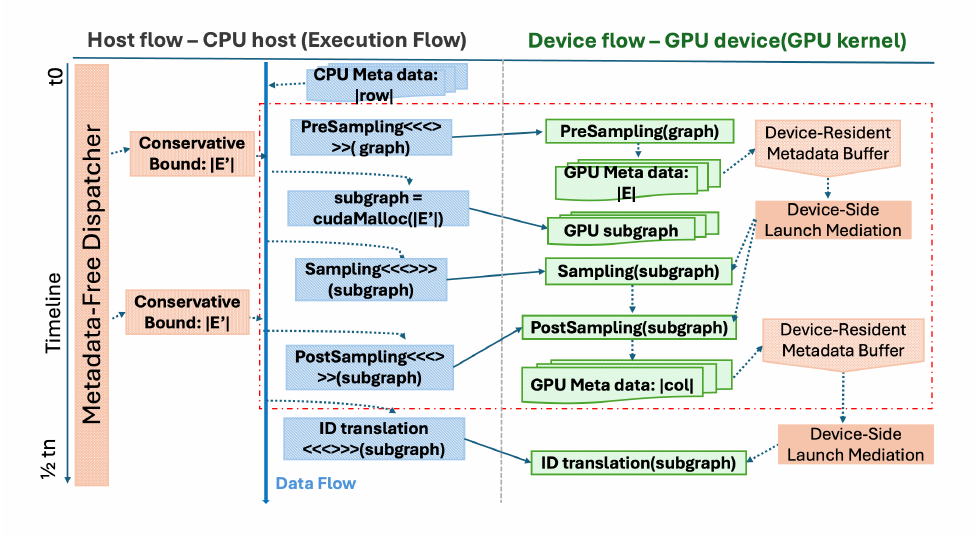}
  \vspace{-12pt}
  \caption{\small {\zerognn} Overview: Device-Resident Metadata-Driven Replayable Execution that removes per-iteration host mediation and improves end-to-end performance.}
  \label{fig-zerognn-overview}
\end{figure}

{\zerognn}'s key idea is restructuring metadata-driven execution so that it remains on-device and becomes replayable. Fig.~\ref{fig-zerognn-overview} illustrates the transformed execution flow.
{\zerognn} uses:
\textbf{1. Device-Resident Metadata Buffer (DRMB).} to keep runtime metadata (e.g., sampled $V_{d}^{N}$ and $E_{d}^{N}$, frontier sizes) in GPU memory and let downstream kernels consume it directly via device pointers, eliminating per-iteration GPU $\rightarrow$ CPU metadata round-trips.
\textbf{2. Device-Side Launch Mediation (DLM)} decouples kernel execution from per-iteration CPU scalar metadata by mediating dynamic work on the GPU: kernels read true runtime sizes from DRMB and specialize execution (e.g., early-exit / bounded access) within a replayable launch structure issued by the host.
\textbf{
3. Metadata-Free Dispatcher (MFD)} makes DLM replayable and safe by provisioning a safe execution envelope, conservative bounds for both memory capacity and launch resources, so the host can issue a fixed launch structure while guaranteeing that device-side specialization in DLM never runs out of bounds, preserving stable addresses and launch shapes for CUDA Graph capture/replay.

Together, these three modules remove per-iteration host mediation in the metadata-driven control loop and restore CUDA Graph replayability under dynamic sampling behavior.

\section{Detailed Design} \label{sec.arch}
This section presents the detailed design of {\zerognn}. Starting from device-resident metadata, we address two follow-on challenges, launch provisioning and memory provisioning, using a replayable execution envelope. We also include two internal baselines only for motivation; they are neither prior work nor adopted by existing systems.

\subsection{Device-Resident Metadata Buffer(DRMB)}
\label{sec.detail.pointer}
The device-resident metadata buffer(DRMB) removes the most fundamental host dependence in sampling-based GNN: runtime metadata is produced on the GPU, but is repeatedly exported to the CPU to drive subsequent execution. DRMB eliminates this export by keeping metadata on-device and making it directly consumable by downstream GPU kernels.

\subsubsection{Design: memory interaction via memory indirection}
DRMB replaces the CPU-visible metadata flow with a GPU-only metadata interaction. Instead of materializing $|V|$ and $|E|$ as CPU scalars, {\zerognn} keeps them resident on the device as GPU pointers, and downstream kernels dereference these pointers to obtain runtime values during execution.
This change turns the metadata dependency from a GPU $\rightarrow$ CPU round-trip into a pure GPU $\rightarrow$ GPU dependency.

Since the number of layers for sampling subgraph objects
is determined by the number of graph convolution layers in
the GNN model, their sizes (i.e., the number of sampling operations and corresponding GPU pointers for their metadata)
are fixed. Hence, memory is allocated once during initial-
ization. Consequently, even if subsequent iterations produce
different vertex and edge counts, the same memory locations
are reused to store the updated values.

This design change enables the CUDA graph
seamlessly: a) the CUDA graph does not support CPU-GPU
synchronization, so removing it with a normal GPU kernel
removes that unsupported code block; and b) a static CUDA
memory pointer is needed for each kernel argument to support
the CUDA graph, which is done using pre-allocating subgraph
metadata.

\subsubsection{Transition: two follow-on provisioning challenges}

With metadata no longer returning to the CPU, two follow-on challenges emerge: (i) how to provision kernel launches when the host cannot access exact runtime sizes, and (ii) how to provision dynamic buffers/tensors without invoking host-side allocators using CPU scalars. We address these challenges next with Device-Side Launch Mediation(\cref{sec.detail.DLM}) and Metadata-Free Dispatcher(\cref{sec.detail.MFD}).

\subsection{Device-Side Launch Mediation(DLM)}
\label{sec.detail.DLM}

We first tackle the launch-side challenge. In metadata-driven GNN training, many kernels require runtime-dependent launch decisions, either computing launch configurations (e.g., grid size from $|V|$ or $|E|$) or determining the number of kernel invocations (e.g., loop-/recursion-based primitives). Since launches are issued from the host, these dependencies would otherwise force metadata back to the CPU and break replayability. Device-Side Launch Mediation(DLM) mediates both cases while preserving a replayable launch skeleton.

\subsubsection{Case 1: Dynamic grid-size allocation}
Kernel launches generally follow either \textit{static allocation} or \textit{dynamic allocation}.
Static allocation launches a fixed grid and relies on grid-stride execution to cover varying input sizes. While it preserves correctness, it is typically an unfavorable fallback for GNN kernels because it often under-utilizes GPU parallelism and weakens tile-per-block optimization structures.

In contrast, dynamic allocation assigns a fixed work quota per block and launches as many blocks as required by the runtime problem size:


\vspace{-10pt}
\begin{equation*}
\text{grid} \approx \left\lceil \frac{N}{T} \right\rceil, 
\quad N \in \{\lvert V \rvert, \lvert E \rvert\}.
\end{equation*}
\vspace{-10pt}

This makes the grid size directly depend on the runtime sizes of $|V|$ and $|E|$. This type of allocation strategy is widely used ~\cite{wang2021gnnadvisor, gespmmsc2020, GNNOne, halfGNN, huang2021understanding, Graphpy} in GNN because it typically yields (i) higher effective occupancy—more blocks than SMs enable better latency hiding; (ii) better load balance—many equal-sized tiles let the hardware scheduler distribute tail work evenly; (iii) improved memory locality and bandwidth efficiency—each block works on a contiguous tile, aiding coalescing and cache reuse; and (iv) easier block-level optimization - fixed-size tiles map naturally to shared-memory buffering, register blocking, and unrolling.

However, once 
$|V|$ and $|E|$ remain on-device via DRMB, the host can no longer compute the exact grid size needed for dynamic allocation without reintroducing GPU $\rightarrow$ CPU synchronization, making this common launch strategy incompatible with replayable execution unless additional mediation mechanisms are applied.

\subsubsection{Case 2: Dynamic number of kernel calls}

Beyond launch dimensions, some computations require invoking kernels multiple times in a loop or recursion, where the number of rounds depends on runtime-dependent sizes such as $|V|$ and $|E|$. A representative example is prefix-sum (scan) used during subgraph construction, e.g., deriving the CSR offset array from the degree array. Scan implementations typically involve two kernels (e.g., scan and add-offset): the first performs a local prefix sum within each CTA, while the second applies CTA-level carry-out offsets. Since the carry-out buffer size itself depends on the input length, one or more additional scan rounds may be required, resulting in a runtime-dependent number of kernel invocations.

This case introduces a stronger form of launch dynamism than Case 1: the host must decide not only the launch configuration, but also how many kernel calls to issue and in what chronological order, which would again require exporting runtime metadata back to the CPU and break replayability.

\subsubsection{Internal baseline: dynamic parallelism via pilot-kernel indirection}
A straightforward baseline to handle both dynamic grid sizing and dynamic kernel-call counts is to shift launch decisions onto the GPU using dynamic parallelism. Since runtime metadata $|V|$ and $|E|$ remain available on-device through DRMB, the system can launch a lightweight pilot kernel whose only purpose is: (i) read the device-resident metadata, (ii) derive the required launch configuration, and (iii) launch actual worker kernels from the device.

This design effectively introduces a launch indirection: the host always launches the pilot kernel with a fixed, capture-friendly configuration (e.g., a single CTA), while the pilot kernel encapsulates the dynamic behavior. Therefore, the host no longer needs to synchronize to fetch $|V|$ and $|E|$, and the pilot kernel can naturally support dynamic control flow such as issuing multiple kernel invocations for loop-/recursion-based primitives (e.g., scan-style operators) until completion.

However, despite being functionally general, this baseline is not practical for sampling-based GNN training. Device-side nested launches introduce substantial overhead and scheduling cost, which directly undermines the goal of reducing host-device orchestration overhead and achieving efficient replayable execution.

\subsubsection{{\zerognn} solution: Device-Side Launch Mediation (DLM)}
DLM avoids nested device-side launches (the internal baseline) and instead mediates launch dynamism with a unified mechanism: the host issues a replayable launch skeleton that is statically capturable, while the GPU specializes execution at runtime using DRMB metadata. Concretely, DLM launches kernels using a conservative upper bound on launch resources (e.g., grid dimensions), provided by our Dispatcher (\cref{sec.detail.MFD}). During execution, each kernel dereferences DRMB to obtain the true runtime sizes (e.g.,  $|V|$ and $|E|$) and applies standard boundary checks so that any over-provisioned blocks exit early without performing out-of-range memory accesses.

\begin{figure}[b]
 \centering
  \includegraphics[scale= 0.55]{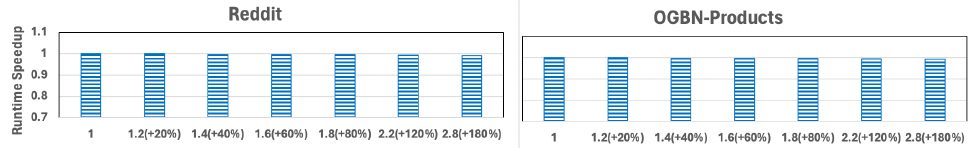}
  \vspace{-12pt}
  \caption{\small SpMM Kernel Runtime Across Different Over-Allocated Grids}
  \label{fig-grid}
  \vspace{-10pt}
\end{figure}

This device-visible metadata is the key enabler: because the exact runtime size is available on-device via DRMB, over-allocation does not compromise correctness—the kernel simply treats the extra blocks as no-ops through early return. In practice, such controlled over-provisioning is also efficient. Fig.~\ref{fig-grid} shows a state-of-the-art SpMM kernel~\cite{GNNOne}  maintains near-constant runtime even when the grid is over-allocated by a large margin (e.g., from +20\% to +180\%) on Reddit and OGBN-Products datasets, indicating that extra blocks can quickly return and incur negligible overhead.

Based on this insight, it is clear that over- or under-allocation
does not prevent the computation from being completed. Further, over-allocation does not introduce a performance regression, while under-allocation may become a minor performance issue if the allocated thread blocks do not saturate the
GPU’s computation power. For this reason, the estimate must
always be higher. However, the estimation must be statically
predictable; otherwise, the estimate itself becomes dynamic
information that cannot solve our original problem of removing dynamic behavior from the CPU execution path.

In summary, DLM eliminates host-visible launch chronology dependence for computations with runtime-dependent control flow (e.g., loop-/recursion-based primitives). Instead of requiring the CPU to decide dynamic launch counts and ordering, DLM keeps the host-side schedule fixed and replayable, while device-resident state mediates whether additional work is required, avoiding metadata export and host-side control dependence.

Finally, to keep launch provisioning statically replayable, the upper bound used by DLM must be fixed during capture and replay. {\zerognn} obtains this bound from the Dispatcher (\cref{sec.detail.MFD}), which provisions a safe execution envelope for both launch and memory resources.

\subsection{Dynamic Memory Provisioning via Dispatch (MFD)}
\label{sec.detail.MFD}

Device-Side Launch Mediation removes GPU $\rightarrow$ CPU metadata export for kernel arguments and enables device-side specialization (\cref{sec.detail.pointer}), dynamic sampling still produces iteration-dependent subgraph sizes. In existing DL frameworks, allocating tensors/buffers (e.g., subgraph structures, feature tensors, intermediate activations) remains a host-side operation that requires CPU-resident scalar sizes such as $|V|$ and $|E|$. If we continue to allocate per iteration based on these runtime sizes, the system must again synchronize and export metadata back to the CPU, effectively reverting back to the HDOO/HMDB bottleneck described in the Introduction (\cref{sec.intro}).
Therefore, enabling replayable execution requires a provisioning strategy that avoids per-iteration host allocation decisions driven by runtime metadata.

Moreover, GNN systems perform tensor allocation through deep learning frameworks (e.g., PyTorch) rather than native CUDA APIs, because memory lifetime management is complex and allocations must be wrapped as framework tensors to support autograd and reuse via caching allocators. However, framework allocation APIs fundamentally require the allocation size to be specified as a CPU-resident scalar. As a result, dynamic tensor provisioning in GNN training—covering sampled subgraphs, associated features and labels, and intermediate activations—forces runtime metadata such as $|V|$ and $|E|$, which are produced on the GPU, to be transferred back to the host. This requirement makes the CPU execution path intrinsically dynamic and introduces unavoidable CPU--GPU synchronization.

\subsubsection{Internal baseline (naïve): Max Reserved Memory}

A straightforward solution is to pre-reserve memory based on the theoretical maximum subgraph size implied by the sampling configuration (i.e., batch size and per-hop fanouts), which we refer to as max reserved memory. This baseline is attractive because it is trivially safe: once the “maximum” buffers are allocated, the pipeline can run without per-iteration metadata queries or GPU $\rightarrow$ CPU synchronization for determining allocation sizes.

However, this baseline is impractical because the maximum subgraph size implied by fan-out sampling grows \emph{multiplicatively} across hops. 
Let $B$ be the mini-batch size, $N$ the number of sampling hops, and $F_i$ the fan-out at hop $i$. 
If each hop expands the frontier by a factor of $F_i$, estimated vertex count after $h$ hops scales as


\vspace{-10pt}
\begin{equation}
V_N \le B \cdot \prod_{i=0}^{N-1} F_i
\end{equation}
\vspace{-9pt}

which is approximately $B \cdot F^{N}$ when $F_i \approx F$, which is exponential in the hop count.
As a result, even moderate fanouts can translate into a highly pessimistic memory budget for sampled node/edge lists, feature tensors, and intermediate activations.

More importantly, this theoretical maximum is rarely reached in real workloads, making the baseline excessively wasteful. First, real-world graphs exhibit strong degree skew, and many nodes simply do not have enough neighbors to satisfy the nominal fan-out, causing systematic sampling shortfall. Second, multi-hop sampling often revisits the same popular neighbors across hops, and deduplication during subgraph construction removes substantial cross-hop duplication. Together, these effects keep the realized subgraph sizes far below the theoretical maximum subgraph size estimate.

Therefore, while max reserved memory eliminates iteration-dependent sizing decisions, it comes at the cost of severe memory over-provisioning, which can easily become infeasible on commodity GPUs and directly limits achievable batch size and model scale. We quantify this effect in our experiments(\cref{sec.exp.memory}).
This motivates the need for a provisioning mechanism that avoids per-iteration host involvement without relying on maximum-capacity reservation.
We next present our solution, the \textit{Metadata-Free Dispatcher (MFD)}, to address this challenge.

\subsubsection{Metadata-Free Dispatcher (MFD): safe-but-tight execution envelope}
\label{sec.detail.MFD.our}

{\zerognn} introduces a \textit{Metadata-Free Dispatcher (MFD)}, which dispatches a conservative yet tight \textit{execution envelope} that can be reused across iterations. The key insight behind MFD is that, in practice, \textit{sampled graph sizes are highly stable across iterations}.

\noindent 
\textbf{Key insight: stability of deduplicated sampled size.}

Although neighbor sampling is stochastic, the resulting computation graph depends only on the deduplicated set of unique sampled vertices, denoted as $V_d$. We find that $|V_d|$ follows a \emph{Poisson-binomial} distribution and, for large graphs, admits a \emph{normal approximation}, which enables a high-confidence concentration bound. We summarize this result as Lemma~\ref{lem:stable_vs}, and provide the full proof in Appendix(\cref{app:mfd_proof}).
We further provide empirical evidence in \cref{sec.exp.dispatch}.

\noindent
\textbf{Formalization.}
To formalize this process, we define an indicator random variable $I_v$ for each vertex $v$, where $I_v = 1$ if $v$ is sampled at least once and $I_v = 0$ otherwise. The deduplicated sampled size is then
\vspace{-10pt}
\begin{equation}
    |V_d| = \sum_{v} I_v .
\end{equation}
\vspace{-10pt}

Under the standard sampling-with-replacement setting, each $I_v$ is a Bernoulli random variable with success probability
\begin{equation}
    p_v = 1 - (1 - \pi_v)^{S_{\text{tot}}},
\end{equation}
where $\pi_v$ can be modeled as a global hitting probability proportional to vertex degree.

Therefore, $|V_d|$ is a sum of non-identical Bernoulli variables, i.e., a \emph{Poisson-binomial} random variable. This distribution concentrates tightly, allowing us to derive the following high-confidence stability bound.

\begin{lemma}[Stability of Deduplicated Sampled Size]
\label{lem:stable_vs}
For multi-hop neighbor sampling with replacement and deduplication, the deduplicated sampled size $|V_d|$ exhibits a narrow, high-confidence fluctuation band around its mean. In particular, for confidence level $p$ over $m$ repeated iterations, the normalized range satisfies:

\vspace{-10pt}
\begin{equation}
\label{eq:vs_range_bound}
    \frac{\max_i |V_d^{(i)}| - \min_i |V_d^{(i)}|}{\mathbb{E}[|V_d|]}
    \le 2 z_{p}^{(m)} \cdot \mathrm{CV},
\end{equation}
\vspace{-10pt}

where $\mathrm{CV} = \sigma/\mu$ is the coefficient of variation of $|V_d|$, and $z_{p}^{(m)}$ is the corresponding Gaussian quantile after accounting for $m$ repetitions.
\end{lemma}

\noindent \textbf{Envelope dispatch.}
Leveraging Lemma~\ref{lem:stable_vs}, MFD dispatches a fixed execution envelope $\mathcal{E}$ that conservatively upper-bounds the resource footprint of dynamic sampling across iterations. Concretely, $\mathcal{E}$ provides (i) \emph{memory provisioning bounds} for sampled-subgraph buffers and scratchpads, and (ii) \emph{launch provisioning bounds} for replayable launch skeleton used by device-side specialization (\cref{sec.detail.DLM}). With $\mathcal{E}$, the host allocates all dynamic buffers once during initialization and reuses the same memory addresses across iterations, preserving, all else equal, the address stability required by CUDA Graph capture/replay. At runtime, kernels dereference DRMB to obtain the true sampled sizes and apply lightweight bounds checks; any excess threads/blocks safely early-exit, ensuring correctness without reintroducing per-iteration host involvement.

\noindent \textbf{Overflow-safe fallback.}
Although Lemma~\ref{lem:stable_vs} enables MFD to dispatch a tight envelope $\mathcal{E}$ with high confidence, it is still a statistical guarantee rather than an absolute bound. Therefore, a vanishingly small fraction of iterations may exceed $\mathcal{E}$ due to rare sampling outcomes. To handle such cases without breaking replayability, {\zerognn} maintains a \emph{cached safe graph} as a backup execution path.

Concretely, during initialization, {\zerognn} caches a \emph{safe graph} (along with its corresponding conservative envelope) that serves as a guaranteed fallback when an overflow is detected. When overflow occurs, the runtime simply replays this cached safe graph, avoiding any per-iteration host-driven reprovisioning. This mechanism can be viewed as maintaining a checkpoint-like backup for execution safety.

Importantly, this fallback does not affect training accuracy. Executing the cached safe graph is semantically equivalent to running the \emph{same computation graph} for the same batch again---i.e., chaining an identical graph twice. This is consistent with the standard full-batch GNN training semantics, where the model update is defined by executing a fixed graph-level computation; thus, the overflow fallback preserves correctness while ensuring memory safety.

\subsection{End-to-End CUDA Graph Capture and Replay}
\label{sec:cuda_graph_replay}

Managing the dynamic behavior of GNN training systematically allows us to keep only predictable control logic on the host (including the Python model code), while removing iteration-by-iteration CPU interventions that primarily serve to orchestrate GPU work. To this end, we leverage \emph{CUDA Graphs} to capture the GPU execution \emph{data-flow} and \emph{kernel launch order} once, and replay it across iterations to eliminate repeated framework overhead.

\noindent \textbf{PyTorch and CUDA Graph.}
CUDA Graph captures a sequence of kernel launches and their dependencies into a graph object. When replaying the same CUDA Graph, the execution does not re-run the original CPU-side launch code; instead, it incurs only a single CUDA Graph launch on the CPU per iteration, while the GPU kernels are issued automatically in the captured order. As a result, each training iteration can be invoked by simply replaying the captured graph, avoiding repeated CPU-side dispatch and launch overhead.

\noindent \textbf{Capture prerequisites and how {\zerognn} restores them.}
CUDA Graph replay relies on a key assumption that each iteration follows the same execution path and operates on tensors with stable shapes and stable memory addresses.
However, metadata-driven dynamic sampling violates this assumption by introducing iteration-dependent provisioning decisions (e.g., buffer sizing and launch configuration), which traditionally require per-iteration host involvement.
{\zerognn} restores the capture/replay conditions by (i) keeping runtime metadata on-device via DRMB (\cref{sec.detail.pointer}), (ii) enabling device-side specialization under a fixed launch skeleton (\cref{sec.detail.DLM}), and (iii) provisioning a replayable execution envelope with stable memory addresses via MFD (\cref{sec.detail.MFD}).

\noindent \textbf{Warm-up, capture, and replay.}
Using CUDA Graph in {\zerognn} follows three phases: \emph{warm-up}, \emph{capture}, and \emph{replay}.
During warm-up, {\zerognn} runs one or more iterations without CUDA Graph to initialize memory pools and materialize required buffers under the MFD-dispatched envelope.
During capture, {\zerognn} records a steady-state training iteration into a CUDA Graph object, including the fixed launch skeleton and all dependent device work.
During replay, {\zerognn} repeatedly replays the captured CUDA Graph for subsequent iterations.

\noindent \textbf{Stable buffers and inputs.}
To ensure replayability, {\zerognn} allocates all dynamic buffers (e.g., sampled-subgraph scratchpads) once and reuses their addresses across iterations (\cref{sec.detail.MFD.our}).
In addition, CUDA Graph replay requires stable input/output tensor pointers.
{\zerognn} therefore performs replay on statically allocated input/output tensors, and copies minibatch features/labels into these fixed buffers before replay.
This ensures that the captured graph remains valid while the training data changes across iterations.

\noindent \textbf{Handling rare overflows.}
Although MFD dispatch is tight with high confidence, extremely rare iterations may exceed the envelope.
{\zerognn} handles such cases by replaying a cached safe graph as a backup execution path (\cref{sec.detail.MFD.our}), and then resuming the steady-state replay loop.
This preserves correctness while keeping the common case fully replayable and host-independent.

\begin{table}[t]
\footnotesize
\caption{\small Dataset details. * indicates labeled dataset, while the rest use 150 generated features and 7 prediction classes.} 
\centering 
\begin{tabular}{l l r r r r} 
    \hline\hline 
    Graph & Vertex & Edge & Feature & Predict\\ [0.5ex] 
    Dataset & Count & Count & Length & Class\\ 
    \hline 
    Cora(G0)* & 2,708 &  10,858 & 1,433 & 7\\ 
    Hollywood(G1) & 1,069,127 & 112,613,308 & 150 & 7\\
    LiveJournal(G2) & 4,847,571 & 137,987,546 & 150 & 7 \\
    OGBN-Products(G3)* & 2,449,029	& 123,718,280 & 100 & 47 \\
    Reddit(G4)* & 232,965 & 229,231,784 & 602 & 41 \\
    Orkut(G5) & 3,072,627 & 234,370,166 & 150 & 7\\
    OGBN-papers100M(G6) & 111,059,956 & 1,615,685,872 & 128 & 172 \\
    \hline 
\end{tabular}
\label{table-dataset} 
\end{table}

\section{Evaluation} \label{sec.exp}

Our evaluation is organized as follows. We first validate accuracy in ~\cref{exp-acc}, then present the main overall results in ~\cref{exp-overall}. We further study large-scale graphs and multi-GPU scaling in ~\cref{exp-large} and in ~\cref{exp-mul-gpu}, followed by a breakdown study in ~\cref{exp-breakdown}. 
Additional supporting results, including a case study against CU-DPI and an analysis of sampled subgraph size variation, are deferred to Appendix~\ref{app:add-exp}.

We compare {\zerognn} with DGL  (version 1.1.0+cu117) ~\cite{dgl2019}, Gong et al.~\cite{Graphpy}, MariusGNN~\cite{waleffe2023mariusgnn}, and CU-DPI, an internal baseline that we define, design, and implement ourselves based on NVIDIA's dynamic parallelism interface to represent a potential alternative for device-side kernel-launch configuration and dynamic memory allocation for runtime metadata. To the best of our knowledge, no existing work or implementation evaluates this design. The datasets are listed in Table~\ref{table-dataset}. Only Cora, Reddit, and OGBN-Products are labeled datasets, and the first two are relatively small. All experiments are conducted on a single node with two NVIDIA A100 GPUs (80GB each) and CUDA 11.7.



\subsection{Accuracy}
\label{exp-acc}
\begin{figure}[b]
 \centering
  \includegraphics[scale= 0.33]{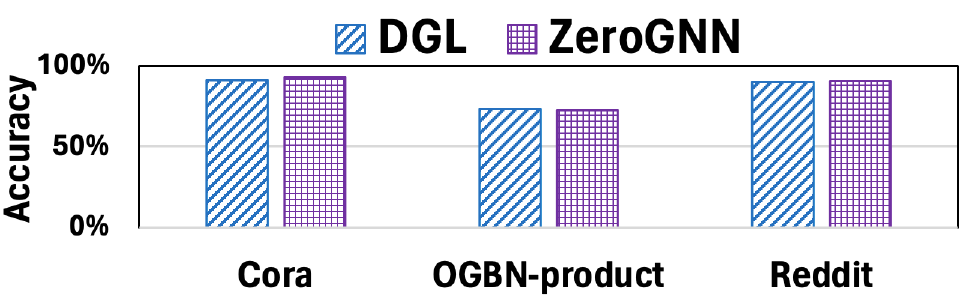}
  \vspace{-12pt}
  \caption{\small Accuracy measure of ZeroGNN and DGL}
  \label{fig-accu}
  \vspace{-19pt}
\end{figure}

We compare {\zerognn} with DGL on the labeled dataset to verify training accuracy. Fig.~\ref{fig-accu} shows that {\zerognn} achieves an accuracy comparable to DGL. 

\subsection{ Overall Evaluation}
\label{exp-overall}
\subsubsection{Performance Speedups}

All subsequent experiments use the GraphSAGE model, following the configuration from Papers~\cite{graphsage17,cluster-gcn}.

\begin{figure}[t]
\begin{minipage}[t]{0.49\columnwidth}
  \includegraphics[width=\linewidth]{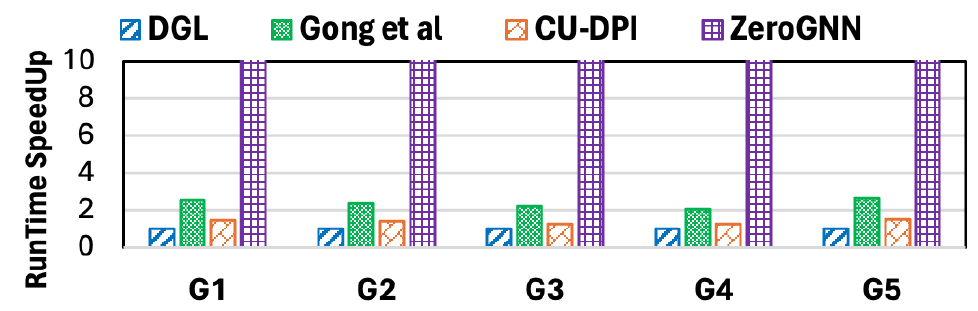}
  \vspace{-16pt}
  \caption{\small Sampling Only Runtime Speedup Over DGL and Gong et al Across Different Datasets. Y-axis Values Are Clipped.
  }
  \label{fig-sample}
\end{minipage} \hfill 
\begin{minipage}[t]{0.48\columnwidth}
  \includegraphics[width=\linewidth]{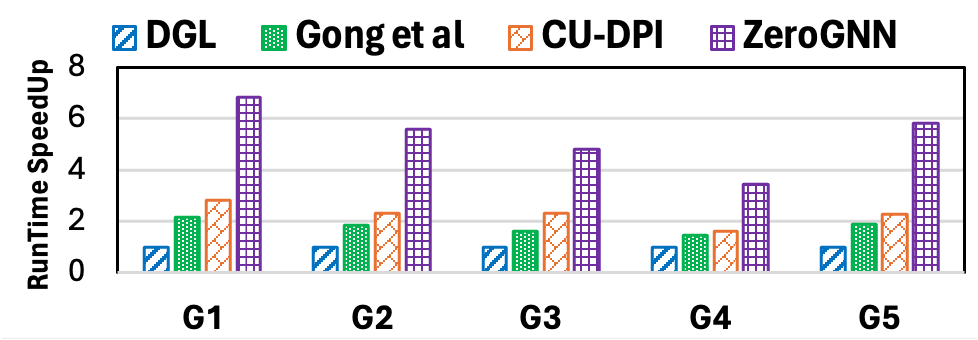}
  \vspace{-16pt}
  \caption{\small End-to-End Runtime Speedup Over DGL Across Different Datasets. 
  }
  \label{fig-all-train-speed}
\end{minipage} 
\end{figure}

\noindent
\textbf{Subgraph Sampling(Stage1)}
Fig.~\ref{fig-sample} shows the comparison of {\zerognn} sampling only runtime with DGL and Gong et al., and CU-DPI across multiple datasets.
For readability, we cap the plotted speedup values to facilitate comparison among baselines.
On average, {\zerognn} achieves a 17.68 $\times$ speedup over DGL.

\noindent
\textbf{End-to-end Training Runtime(All Stages)}
Fig.~\ref{fig-all-train-speed} shows the comparison of {\zerognn} end-to-end training runtime with DGL, Gong et al. and CU-DPI across multiple datasets. On average, ZeroGNN achieves a 5.28 $\times$ speedup over DGL.


\subsubsection{Memory Usage Analysis}
\label{sec.exp.memory}

We compare the memory usage of {\zerognn} with DGL, Gong et al., and MaxSG. Gong et al. serves as an optimal reference because it allocates memory using exact runtime metadata, while MaxSG represents a naive worst-case reservation strategy.

\begin{figure}[b]
 \centering
  \includegraphics[scale= 0.45]{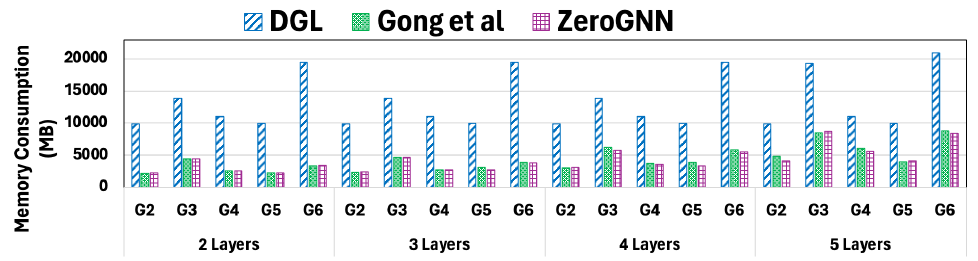}
  \vspace{-12pt}
  \caption{\small Memory usage comparison between {\zerognn}, DGL, and the optimal dynamic allocation baseline(Gong et al) across different sampling depths(L2–L5; lower is better).
}
  \label{fig-memory-all}
  \vspace{-32pt}
\end{figure}

\noindent
\textbf{Memory Usage} Fig.\ref{fig-memory-all} shows that {\zerognn} achieves a memory footprint comparable to this optimal, dynamically informed solution (Gong et al), while significantly outperforming DGL. Notably, both DGL and Gong et al rely on dynamic metadata and adaptive allocation, whereas {\zerognn} attains near-optimal memory efficiency through device-side indirection and fixed subgraph sampling mechanisms—without requiring per-iteration metadata updates.

\begin{figure}[t]
 \centering
  \includegraphics[scale= 0.43]{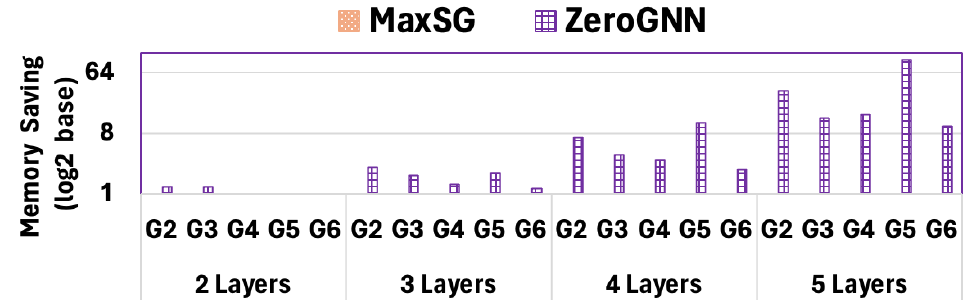}
  \vspace{-12pt}
  \caption{\small Memory usage efficiency comparison between {\zerognn} and the MaxSG (Naive Maximum Subgraph Allocation strategy) across different sampling depths. The MaxSG serves as the baseline (value = 1). Efficiency is measured on a $\log_{2}$ scale, where higher values indicate better memory efficiency.
}
  \label{fig-memory-maxg}
\end{figure}

\noindent
\textbf{Comparison with MaxSG(Internal Baseline)}
We further compare {\zerognn} with MaxSG, a naive allocation strategy, which always reserves the maximum possible memory for subgraph sampling based solely on the batch size—ignoring the actual subgraph size in each iteration. The experimental results, presented in Fig.\ref{fig-memory-maxg}, reveal distinct memory utilization trends across different GNN layers.
On average, {\zerognn} delivers a 10.84 $\times$ memory efficiency improvement over the MaxSG baseline.

In layer 2, the subgraph size remains small, leading to only marginal differences between {\zerognn} and the naive approach. However, as we progress to layers 3–5, the advantages of {\zerognn} become increasingly evident. This improvement arises because deeper layers generate substantially larger subgraphs with more duplicated or overlapping neighborhoods. {\zerognn}’s optimized allocation strategy mitigates such redundancy via metadata reuse and fine-grained device-side allocation, while the naive solution continues to over-provision memory for the worst-case scenario.

For clarity, the results are shown on a $\log_{2}$ scale, where the naive allocation(MaxSG) is normalized to 1, and the {\zerognn} bars represent the $\log_{2}$-based memory savings relative to this baseline. As observed, {\zerognn} consistently achieves significant memory reductions, especially in deeper layers where dynamic reuse opportunities are amplified.

\subsection{Large-scale Graph}
\label{exp-large}

Large-scale graphs such as OGBN-papers100M often exceed GPU memory capacity primarily because of the full node feature table rather than the graph topology itself. In our setting, the graph topology is placed on the GPU to enable device-side subgraph sampling, but storing the full feature table on the GPU leads to out-of-memory errors. To demonstrate that {\zerognn} remains effective under such large-graph conditions, we construct a simulated large-graph setting by replacing the full feature table with a GPU-resident feature buffer while keeping the graph topology and GPU-side sampling path unchanged. This remains representative because sampled GNN training is performed on sampled subgraphs rather than on the full graph: once features for the sampled nodes are supplied, the downstream training computation remains unchanged. 
In this sense, the simulation removes only the full-table memory bottleneck, without changing the sampled-subgraph sampling and training path that we aim to evaluate.

\begin{figure}[b]
 \centering
  \includegraphics[scale= 0.38]{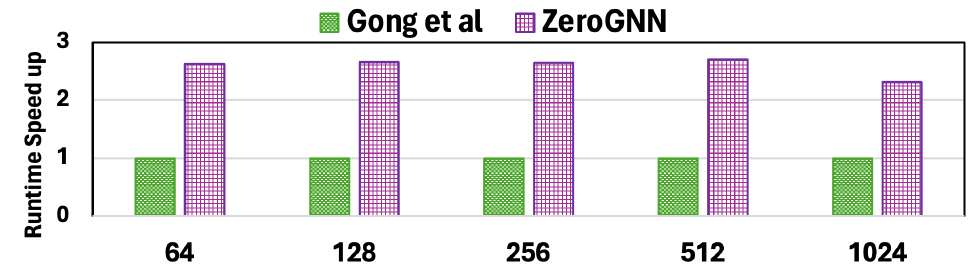}
  \vspace{-12pt}
  \caption{\small End-to-End Runtime Speedup For {\zerognn} Over Gong et al Across Different Batch Sizes On OGBN-papers100M Dataset. }
  \label{fig-large-graph}
  \vspace{-13pt}
\end{figure}

Under this setting, DGL still runs out of memory because of additional framework overheads, while MariusGNN~\cite{waleffe2023mariusgnn} is incompatible with our fully GPU-based sampling configuration due to its tightly coupled disk/CPU-driven pipeline. We therefore compare against Gong et al. Fig.~\ref{fig-large-graph} shows that {\zerognn} achieves 2.31$\times$ to 2.70$\times$ performance speedups over Gong et al across different batch sizes.

\subsection{Multi-GPU Scaling}
\label{exp-mul-gpu}
In this section, we show that {\zerognn} remains effective under multi-GPU training. Multi-GPU training follows a data-parallel execution model~\cite{p3gandhi2021, torch-data-parallel, Horovod, distributed-para}, in which each GPU processes a distinct mini-batch on its own model replica, and gradients are synchronized across devices after each iteration. However, scaling across multiple GPUs does not eliminate the core inefficiency of sampling-based GNN execution. Within each worker, subgraphs are still generated dynamically, runtime metadata still determines downstream tensor allocation and kernel launches, and host-side orchestration remains on the critical path in existing systems. Although multi-GPU training introduces inter-GPU communication(e.g., via all-reduce), this communication is not the primary bottleneck, since the communicated data mainly consists of model parameters, which are relatively small in sampling-based GNN models. As a result, even in the multi-GPU setting, the dominant limitation still lies in the per-worker, metadata-driven host control path rather than in gradient synchronization.

\begin{figure}[t]
\begin{minipage}[t]{0.49\columnwidth}
  \includegraphics[width=\linewidth]{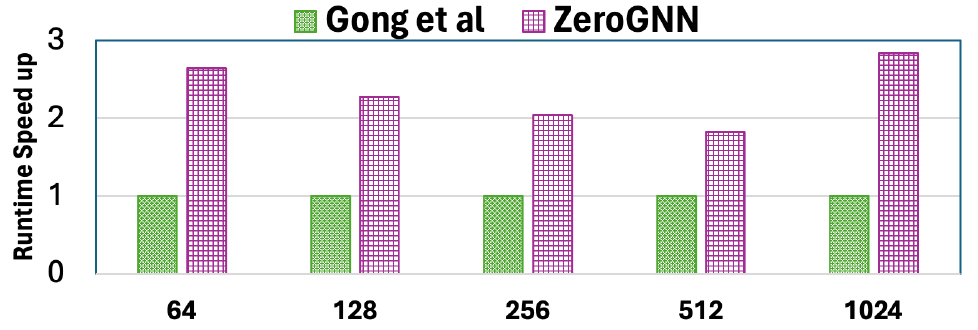}
  \vspace{-22pt}
  \caption{\small {\zerognn} End-to-End Training Runtime Comparison Under 2 GPUs Configurations.
  }
  \label{figure-multi-gpu-speed-up}
\end{minipage} \hfill 
\begin{minipage}[t]{0.49\columnwidth}
  \includegraphics[width=\linewidth]{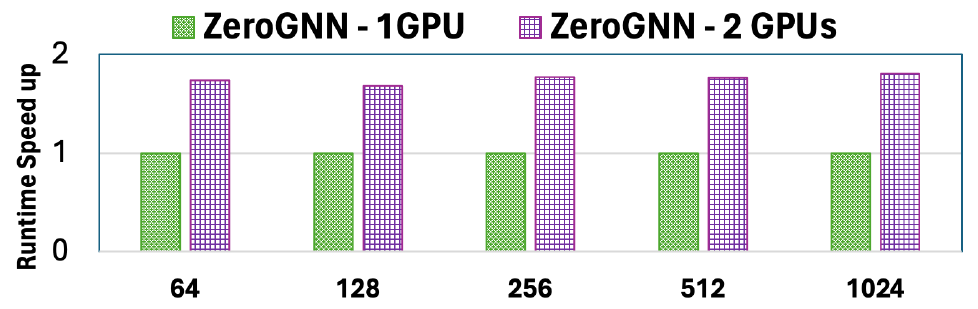}
  \vspace{-22pt}
  \caption{\small {\zerognn} End-to-End Training Runtime Comparison for strong scaling. 
  }
  \label{figure-multi-gpu-scale}
\end{minipage} 
\end{figure}

Fig.~\ref{figure-multi-gpu-speed-up} compares {\zerognn} with Gong et al on a 2-GPU machine and shows that {\zerognn} achieves an average speedup of up to 8$\times$ in the multi-GPU setting. This gain arises because the baseline still executes through a GPU$\rightarrow$CPU$\rightarrow$GPU control path: runtime metadata must be transferred to the host, used to determine subsequent tensor provisioning and kernel launch decisions, and then translated back into GPU work. Such repeated host involvement introduces persistent synchronization, launch preparation, and orchestration overhead that is not removed by multi-GPU execution. As a result, {\zerognn} continues to deliver strong end-to-end speedups over the baseline across different batch sizes.

Fig.~\ref{figure-multi-gpu-scale} shows the 1-to-2 GPU strong-scaling results of {\zerognn}. Across batch sizes, {\zerognn} achieves 1.68$\times$ – 1.80$\times$ speedup, indicating consistently strong scaling with only modest variation. This stability is important because, at smaller batch sizes, framework overhead would normally account for a larger fraction of end-to-end execution time, making speedup less stable and causing performance to approach the ideal 2$\times$ only at larger batch sizes. Instead, {\zerognn} maintains similar scaling behavior across the entire range, indicating that the framework overhead that would otherwise disproportionately penalize small-batch execution has been effectively removed from the critical path. Overall, {\zerognn} not only preserves its end-to-end benefit under multi-GPU training, but also delivers stable strong scaling across workloads.

\subsection{Speedup Analysis and Breakdown}
\label{exp-breakdown}
\subsubsection{GPU Execution Fraction}

This section evaluates GPU Execution Fraction between {\zerognn} and existing systems. We demonstrate that {\zerognn} achieves substantially higher utilization than prior work, from the sampling perspective as well as the end-to-end training perspective. We omit CU-DPI from the plot, as the dynamic parallelism interface disables GPU time profiling.~\cite{dp_no_profile}

\begin{figure}[t]
\begin{minipage}[t]{0.49\columnwidth}
  \includegraphics[width=\linewidth]{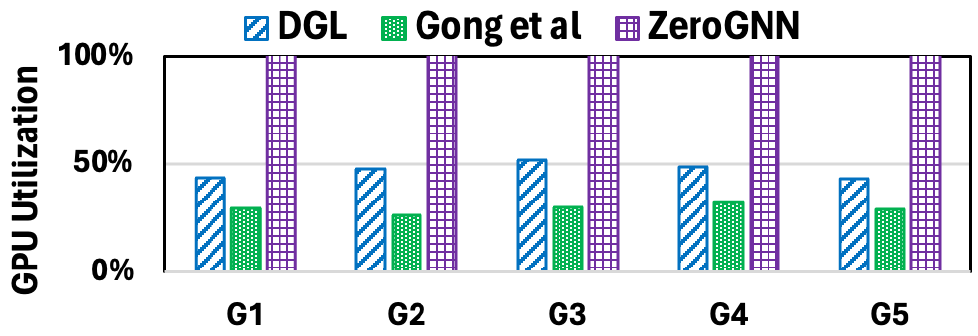}
  \vspace{-22pt}
  \caption{\small Comparison of GPU Execution Fraction across Systems during the Sampling Stage.
  }
  \label{fig-gpu-util-sample}
\end{minipage} \hfill 
\begin{minipage}[t]{0.49\columnwidth}
  \includegraphics[width=\linewidth]{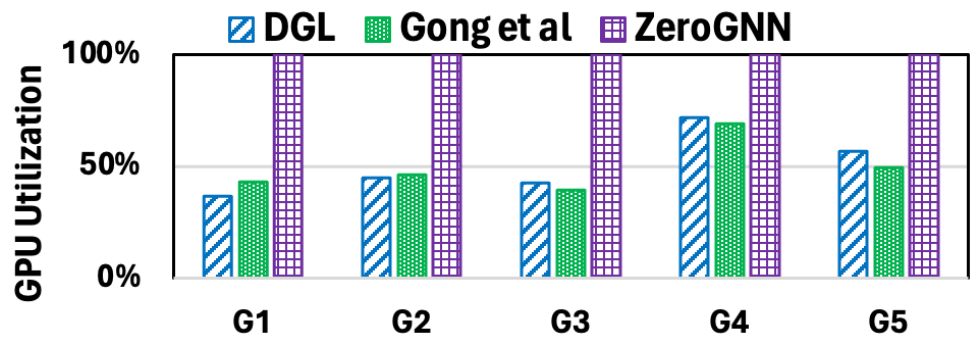}
  \vspace{-22pt}
  \caption{\small Comparison of GPU Execution Fraction across Systems during the end-to-end training. 
  }
  \label{fig-gpu-util-all}
\end{minipage} 
\end{figure}

\noindent
\textbf{Subgraph Sampling(Stage1)}
Fig.~\ref{fig-gpu-util-sample} reports GPU Execution Fraction in the sampling stage. {\zerognn} sustains ~100\% utilization, indicating negligible kernel idle time. In contrast, Gong et al. and DGL show substantially lower utilization (with Gong et al below DGL), demonstrating that kernel-level tweaks alone cannot overcome persistent framework overheads that throttle end-to-end utilization.

\noindent
\textbf{End-to-end Training(All Stages)}
Fig.~\ref{fig-gpu-util-all} shows end-to-end GPU Execution Fraction across systems. ZeroGNN maintains ~100\% utilization, whereas existing baselines exhibit substantially lower GPU execution fraction. This indicates that our approach effectively removes framework overhead, resolving the dominant bottleneck.

\subsubsection{End-to-End Performance Scaling Analysis}
In this section, we examine how two additional input factors, batch size and sampling depth, affect the end-to-end performance speedup of {\zerognn}.

\begin{figure}[t]
 \centering
  \includegraphics[scale= 0.37]{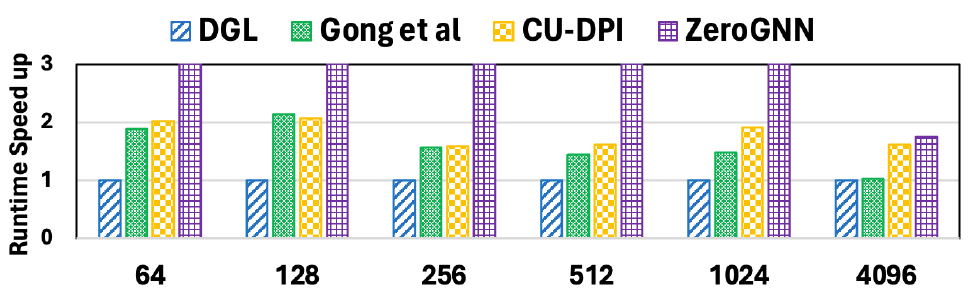}
  \vspace{-12pt}
  \caption{\small End-to-End Runtime Speedup For {\zerognn} Over DGL Across Different Batch Sizes On Reddit Dataset.}
  \label{fig-batch}
\end{figure}

\noindent
\textbf{Batch Size}
We fix all other configurations (e.g., fanout and layouts) and vary only the batch size. 
Fig.~\ref{fig-batch} shows the end-to-end runtime speedup of {\zerognn} over existing systems on Reddit. 
{\zerognn} achieves, on average, a 4.80 $\times$(2.84 $\times$) performance speedup compared to DGL(Gong el al).
For readability, the y-axis is clipped at 3$\times$, so bars reaching the top correspond to speedups of at least 3$\times$.
For example, at batch size 64, {\zerognn} achieves a 8.75 $\times$ speedup compared to DGL.
It is clear that when the batch size is small, such as 64, {\zerognn} achieves much better performance speedups than all existing works.
As batch size increases, the speedup decreases, because GPU compute accounts for a larger fraction of end-to-end runtime, whereas {\zerognn} mainly removes framework overhead rather than optimizing GPU kernels. 
Even at batch size 4096, {\zerognn} still achieves a 1.75 $\times$ speedup, while the existing work Gong et al attains only a runtime similar to DGL.

\begin{figure}[t]
 \centering
  \includegraphics[scale= 0.35]{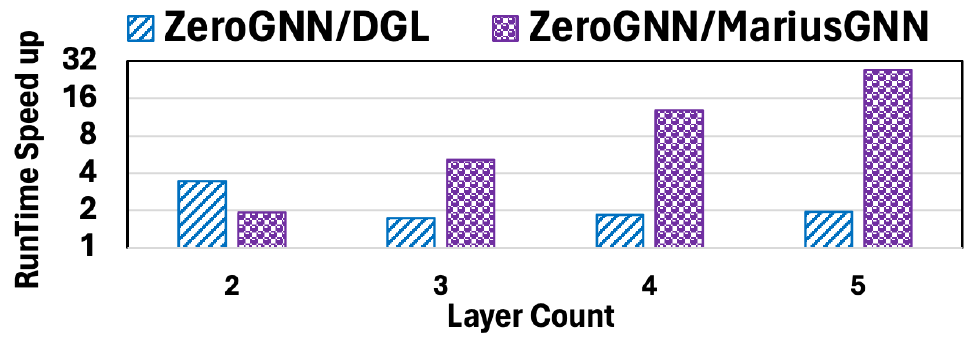}
  \vspace{-14pt}
  \caption{\small End-to-End Runtime Speedups (Log$_2$ Scale) across Sampling Layers on the Reddit Dataset.}
  \label{fig-multi-hop}
\end{figure}

\noindent
\textbf{Sampling Model Depths: Multi-Hop Sampling}
Fig.~\ref{fig-multi-hop} shows that {\zerognn} achieves the largest speedups at smaller model depths, while the benefit decreases moderately as depth increases but remains higher than all baselines. Specifically, {\zerognn} achieves a 3.50$\times$ speedup at 2 layers and still delivers 1.96$\times$ at 5 layers. This trend arises because deeper models generate larger sampled subgraphs, increasing GPU compute time while framework overhead remains roughly constant. Since {\zerognn} primarily removes framework overhead, its relative benefit decreases as GPU computation becomes more dominant, though it remains consistently beneficial across all depths.

\section{Related Work} \label{sec.related}


\noindent
\textbf{Optimizing Sampling-Based GNN Systems}
Prior works improve the efficiency of sampling-based GNN execution by optimizing sampling, pipeline organization, or data placement. SALIENT ~\cite{kaler2022accelerating} improves sampling efficiency mainly from the CPU/framework side through a lightweight C++ implementation and reduced software overhead, while NextDoor ~\cite{jangda2021accelerating} accelerates GPU sampling through frontier-aware execution and better load balancing. MariusGNN~\cite{waleffe2023mariusgnn} improves subgraph construction efficiency through partitioning and locality-aware design. Other systems, such as GNNLab ~\cite{yang2022gnnlab}, DSP ~\cite{DSP23}, and WholeGraph ~\cite{Yang2022WholeGraphAF}, improve end-to-end throughput through pipeline restructuring, cache/locality optimization, multi-GPU partitioning, and communication-efficient data access. These works improve sampling or system throughput, but they do not eliminate the metadata-driven host control path that determines downstream provisioning and kernel launch behavior within each iteration.

\noindent
\textbf{GNN computation and kernel optimization.}
Another line of work focuses on optimizing GPU kernel computation itself. Prior studies, including DA-SpMM~\cite{DA-SpMM}, GE-SpMM~\cite{gespmmsc2020}, GNNOne~\cite{GNNOne}, and related operator-level optimizations~\cite{fu2022tlpgnn, gale2020sparse,huang2021understanding, wu2021seastar, yang2018design, featgraphsc20,chen2020fusegnn}, improve sparse-dense computation through input-aware kernel design and execution specialization. HalfGNN~\cite{halfGNN} further improves training efficiency under half precision. These works target the efficiency of GPU kernels operating on the sampled graph, whereas our work addresses a different bottleneck: the host-side orchestration overhead caused by runtime metadata across the sampling-to-training pipeline. 

\noindent
\textbf{CPU–GPU Communication Optimization}
Prior works ~\cite{CGCM, DyManD, pai_cpugpu_commu, li_cpugpu_commu} also reduced CPU–GPU interaction overhead through communication and memory-management optimizations. Examples include compiler/runtime techniques that remove redundant transfers, demand-driven migration for pointer-based workloads, and coherence- or placement-based approaches that keep data near the processor that uses it. These methods are effective when the goal is to reduce redundant communication or exploit persistent data reuse across accesses. In contrast, our setting is fundamentally different: it is driven by iteration-specific runtime metadata rather than reusable data across accesses, and thus concerns the placement of such metadata in a host-mediated control loop rather than the reuse, retention, or migration of persistent data.

\section{Conclusion and Future Works} \label{sec.conclusion}

We analyze sampling-based GNN systems and observe that their training runtime is dominated by framework overhead rather than GPU computation. However, the dynamic metadata workflow generated in each iteration repeatedly forces host-mediated coordination, introducing synchronization into the critical path and preventing CUDA Graph replay from effectively removing this overhead. To address this gap, we propose {\zerognn}, which combines device-resident metadata, device-side specialization, and a conservative replayable execution envelope to enable dynamic execution under a fixed host launch structure. Our results show that this design substantially improves end-to-end performance and scalability. More broadly, the same perspective may be useful for other metadata-driven dynamic workloads beyond GNN training.

\newpage
\clearpage
 \balance
 {
 \bibliographystyle{refer/abbrv}
 \bibliography{refer/GraphPy}
 }

\newpage

\appendix

\section{Proof of Lemma~\ref{lem:stable_vs}}
\label{app:mfd_proof}

\paragraph{Problem setting.}
We consider the standard multi-hop neighbor sampling procedure widely used in GraphSAGE-like GNN training~\cite{graphsage17}.
Given an input mini-batch of seed nodes $S^{(0)}$ (batch size $B$), the sampling expands hop by hop.
For each hop $h=1,2,\ldots,H$, we are given the current frontier $S^{(h-1)}$, and for each vertex $u \in S^{(h-1)}$, we sample its neighbors \emph{with replacement} (uniformly among all neighbors).
All sampled neighbors are collected into $S^{(h)}$.
After $H$ hops, the raw sampled node multiset is:
\begin{equation}
S^{(0)} \cup S^{(1)} \cup \cdots \cup S^{(H)}.
\end{equation}
Although sampling is performed with replacement, the final computation graph used by the GNN depends only on the set of \emph{unique} sampled nodes.
Thus, we define the deduplicated sampled vertex set as
\begin{equation}
V_s = \mathrm{unique}\!\left(S^{(0)} \cup S^{(1)} \cup \cdots \cup S^{(H)}\right),
\end{equation}
and the quantity we care about is the deduplicated sampled size $|V_s|$.

\paragraph{Goal of the proof.}
In typical GNN training, this multi-hop sampling procedure is executed repeatedly once per iteration (or per mini-batch), producing a sequence:
\begin{equation}
|V_s^{(1)}|,\ |V_s^{(2)}|,\ \ldots,\ |V_s^{(m)}|.
\end{equation}
We aim to show that even though sampling is random and $|V_s|$ fluctuates from iteration to iteration, its fluctuation range is extremely small when expressed as a percentage of its mean:
\begin{equation}
\frac{\max_i |V_s^{(i)}| - \min_i |V_s^{(i)}|}{\mathbb{E}[|V_s|]} \ll 1,
\end{equation}
with very high confidence (e.g., $99.9\%$ or $99.99\%$).

\paragraph{Modeling a draw via a global hitting probability.}
We model each neighbor-sampling draw as selecting one element from the multiset of \emph{neighbor-occurrences}.
When we take the union of all neighbor-occurrences across all seeds and hops, we obtain a large pool.
Hence, each draw can be viewed as choosing one element from this global pool.
Consequently, the probability that a sampling draw returns vertex $v$ can be modeled by the global hitting probability:
\begin{equation}
\pi_v = \frac{\deg(v)}{\sum_{u\in V} \deg(u)}.
\end{equation}
This definition ensures that $\pi_v \ll 1$ and typically $\pi_v S_{\mathrm{tot}} \ll 1$, which matches the regime where the Poisson and Poisson-binomial approximations are accurate.

\paragraph{Formalization with indicator random variables.}
We define an indicator random variable for each vertex $v$:
\begin{equation}
I_v=
\begin{cases}
1, & \text{if vertex $v$ appears in the sample at least once},\\
0, & \text{otherwise}.
\end{cases}
\end{equation}
Then the deduplicated total node count is
\begin{equation}
|V_s| = \sum_{v=1}^{n} I_v,
\end{equation}
where $n$ is the total number of vertices in the graph.

\paragraph{Probability that a vertex gets sampled.}
Assume the sampling pipeline performs a total of $S_{\mathrm{tot}}$ draws, and each draw hits vertex $v$ with probability $\pi_v$.
Then the probability that vertex $v$ is hit at least once is:
\begin{equation}
p_v = 1-(1-\pi_v)^{S_{\mathrm{tot}}}.
\end{equation}
When $\pi_v$ is small and $S_{\mathrm{tot}}$ is large (as sampling-based GNNs operate on sparse, dynamically sampled subgraphs),

\begin{equation}
(1-\pi_v)^{S_{\mathrm{tot}}} \approx e^{-\lambda_v},
\qquad
\lambda_v = S_{\mathrm{tot}}\pi_v,
\end{equation}
thus
\begin{equation}
p_v \approx 1-e^{-\lambda_v}.
\end{equation}
Therefore,
\begin{equation}
I_v \sim \mathrm{Bernoulli}(p_v).
\end{equation}

\paragraph{Overall distribution: Poisson-binomial.}
Because the indicators $I_v$ are (approximately) independent Bernoulli random variables with different success probabilities $p_v$, the total number of unique sampled nodes is
\begin{equation}
|V_s| = \sum_{v=1}^{n} I_v,
\end{equation}
which follows a Poisson-binomial distribution:
\begin{equation}
|V_s| \sim \mathrm{PB}(p_1,p_2,\ldots,p_n).
\end{equation}

\paragraph{Approximate normal distribution.}
When the number of vertices $n$ is large and the sampling probabilities $p_v$ are not extreme, the central limit theorem for Poisson-binomial applies:
\begin{equation}
\frac{|V_s|-\mu}{\sigma} \Rightarrow \mathcal{N}(0,1),
\end{equation}
where
\begin{equation}
\mu=\sum_v p_v,
\qquad
\sigma^2=\sum_v p_v(1-p_v).
\end{equation}
Therefore,
\begin{equation}
|V_s| \approx \mathcal{N}(\mu,\sigma^2).
\end{equation}

\paragraph{Range (fluctuation) of sampled node count across $m$ iterations.}
Since $|V_s|$ is approximately normal, we can compute its confidence interval.
For a given confidence level $p$ and sampling repeated $m$ times, define:
\begin{equation}
z_p^{(m)} = \Phi^{-1}\!\left(p^{1/m}\right),
\end{equation}
where $\Phi^{-1}$ is the inverse CDF of the standard normal distribution.
Then the confidence interval for all $m$ repeated samplings is:
\begin{equation}
[T_{\min},T_{\max}] =
\left[\mu - z_p^{(m)}\sigma,\ \mu + z_p^{(m)}\sigma\right].
\end{equation}
Thus, the range is
\begin{equation}
\mathrm{Range} = T_{\max}-T_{\min} = 2z_p^{(m)}\sigma.
\end{equation}

\paragraph{Normalized range and coefficient of variation.}
To compare fluctuations across different sampling settings, we normalize by the mean:
\begin{equation}
r_{\mathrm{range}}
=
\frac{\mathrm{Range}}{\mu}
=
\frac{2z_p^{(m)}\sigma}{\mu}
=
2z_p^{(m)}\cdot \mathrm{CV},
\end{equation}
where
\begin{equation}
\mathrm{CV}=\frac{\sigma}{\mu}
\end{equation}
is the coefficient of variation.

\paragraph{Core insight: $\mathrm{CV}$ depends only on $p_v$.}
Recall:
\begin{equation}
|V_s|=\sum_v I_v,
\qquad
I_v\sim\mathrm{Bernoulli}(p_v).
\end{equation}
Thus,
\begin{equation}
\mu=\sum_v p_v,
\qquad
\sigma^2=\sum_v p_v(1-p_v),
\end{equation}
and the coefficient of variation is:
\begin{equation}
\mathrm{CV}
=
\frac{\sigma}{\mu}
=
\sqrt{\frac{\sum_v p_v(1-p_v)}{\left(\sum_v p_v\right)^2}}.
\end{equation}

\paragraph{Sparse-sampling regime: all $p_v$ very small.}
When $p_v$ is very small, $p_v(1-p_v)\approx p_v$, hence
\begin{equation}
\sigma^2 \approx \sum_v p_v = \mu,
\qquad
\mathrm{CV}\approx \frac{1}{\sqrt{\mu}}.
\end{equation}
Therefore, larger sampling budgets lead to more stable sampled graph sizes.

\paragraph{Final conclusion.}
Combining the above steps, the sampling stability satisfies:
\begin{equation}
r_{\mathrm{range}} = 2z_p^{(m)}\cdot \mathrm{CV},
\end{equation}
which concludes the proof.

\section{Additional Evaluation Results}
\label{app:add-exp}

\subsection{Case Study: Why ZeroGNN Outperforms the CU-DPI Baseline}
\label{exp-case}
\noindent
\textbf{CU-DPI: CUDA Dynamic-Parallelism Implementation.}
In this design, a pilot kernel dynamically launches child kernels on demand. Although this approach simplifies programming, it introduces significant kernel-launch overhead, since each child kernel invocation behaves as an independent launch.

\noindent
\textbf{ZeroGNN: Over-provisioned launch with device-side indirection.}
{\zerognn} launches kernels using an estimated upper bound for grid size and memory. Inside the kernel, the true work sizes are read from GPU memory via GPU pointer (indirection), so only the necessary blocks perform work, and allocated buffers are sized effectively for the actual metadata.

\begin{figure}[t]
 \centering
  \includegraphics[scale= 0.45]{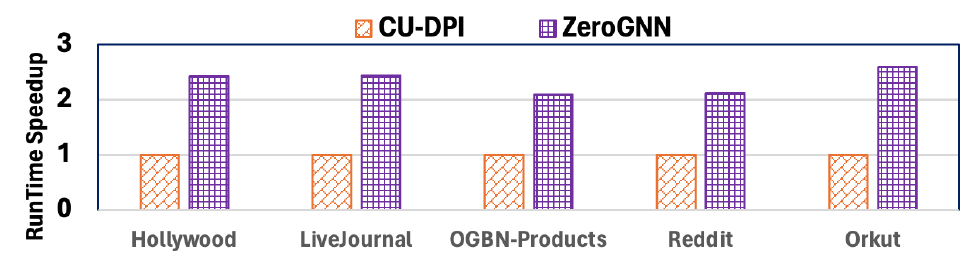}
  \vspace{-12pt}
  \caption{\small End-to-End Training Runtime Comparison Across Different Design Choices to handle  Dynamic Dataflow.}
  \label{figure-zerognn-design-choice}
\end{figure}

Our experiments show that CU-DPI introduces noticeable overhead due to repeated kernel launches, whereas {\zerognn} effectively eliminates this cost. Fig.~\ref{figure-zerognn-design-choice} quantifies the overhead of CU-DPI relative to our {\zerognn} design.

\subsection{Sampled Subgraph Size Variation Analysis} 
\label{sec.exp.dispatch}

\begin{figure}[b]
 \centering
  \includegraphics[scale= 0.5]{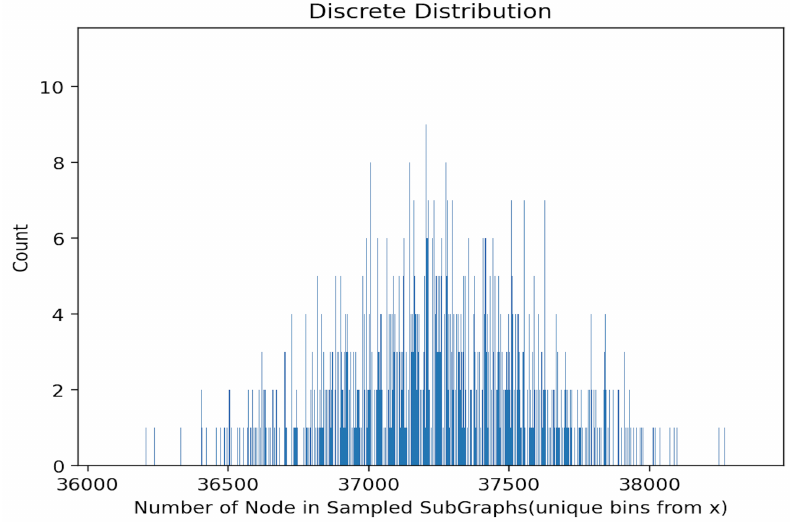}
  \vspace{-12pt}
  \caption{\small Distribution of sampled subgraph sizes (number of nodes) for {\zerognn} on the Reddit dataset.}
  \label{fig-upper-bound}
\end{figure}

In this section, we analyze the distribution of sampled subgraph sizes to demonstrate that their variability is minimal. We perform the experiments by using the default model configuration on the Reddit dataset. 
Fig.~\ref{fig-upper-bound} shows the experiment result. 
The size (number of nodes) of the subgraph obtained at each training iteration is plotted on the X axis. The resulting histogram is approximately bell-shaped: most samples cluster near the center, and only a few outliers appear. Quantitatively, the percentage spread between the observed maximum and minimum subgraph sizes is ~7\%, which is well below the 20\% upper-bound margin we provisioned. This indicates that across iterations, sampled subgraph sizes remain tightly concentrated and do not fluctuate substantially.

\end{document}